\documentclass[manuscript,screen]{acmart}
\usepackage {graphicx}
\usepackage{subfigure}
\usepackage{biblatex}
\usepackage{wrapfig}
\usepackage{geometry}
\usepackage{pdflscape}
\usepackage{multirow}
\usepackage{ragged2e}
\usepackage{rotating}

\AtBeginDocument{%
  \providecommand\BibTeX{{%
    \normalfont B\kern-0.5em{\scshape i\kern-0.25em b}\kern-0.8em\TeX}}}

\setcopyright{acmcopyright}
\copyrightyear{2023}
\acmYear{2023}
\acmDOI{10.1145/1122445.1122456}

\begin{document}

\title{A Survey on Acoustic Side Channel Attacks on Keyboards}

\author{Alireza Taheritajar}
\email{ataheritajar@augusta.edu}
\orcid{0009-0002-5642-6329}
\affiliation{%
  \institution{Augusta University}
  \city{Augusta}
  \state{Georgia}
  \country{USA}
}

\author{Zahra Mahmoudpour Harris}
\email{zmahp65@gmail.com}
\affiliation{%
  \institution{Shariaty Technical College}
  \city{Tehran}
  \state{Tehran}
  \country{Iran}
}

\author{Reza Rahaeimehr}
\email{rrahaeimehr@augusta.edu}
\orcid{0000-0003-0305-3661}
\affiliation{%
  \institution{Augusta University}
  \city{Augusta}
  \country{USA}}
\email{rrahaeimehr@augusta.edu}

\renewcommand{\shortauthors}{Alireza Taheritajar et al.}

\begin{abstract}
Most electronic devices utilize mechanical keyboards to receive inputs, including sensitive information such as authentication credentials, personal and private data, emails, plans, etc. However, these systems are susceptible to acoustic side-channel attacks. Researchers have successfully developed methods that can extract typed keystrokes from ambient noise.
As the prevalence of keyboard-based input systems continues to expand across various computing platforms, and with the improvement of microphone technology, the potential vulnerability to acoustic side-channel attacks also increases. This survey paper thoroughly reviews existing research, explaining why such attacks are feasible, the applicable threat models, and the methodologies employed to launch and enhance these attacks.
\end{abstract}



\maketitle
\section{Introduction}

With emerging cyber crimes in the 1980s, hacking and stealing data, and consequently, data privacy and confidentiality, became the new concerns for computer scientists. Since then, scientists have tried to find better mechanisms to secure systems, and hackers have attempted to break the current mechanisms or discover new vulnerabilities. Although several well-thought privacy and confidentiality preserving mechanisms have been introduced in theory, achieving a perfectly secure system is impossible in practice. One reason is that every process in the world has some side effects; Every process consumes some energy and may have noise, generate vibration, emit optical particles, change electromagnetic fields, produce heat, and delay the execution of other processes. And all these side effects, to some extent, leak information about the process.

In computer science, using the side effects of a process to gain knowledge about the process is called Side-Channel Attack. This type of attack is applicable on any system, no matter how carefully the system is designed.  
In 1973, Butler W. Lampson et al.\cite{lampson1973note} discussed the possibility of abusing intermediate files and resources used or generated during a process. In 1985, a research project\cite{van1985electromagnetic} explained the possibility of eavesdropping on display units by gathering and analyzing their electromagnetic interference. In 1996, Paul Kocher et al.\cite{kocher1996timing} published the first significant attack in this area. They discovered they could break some RSA-based cryptosystems by analyzing the time it takes to do the private key operations. The attacks \cite{Lipp2018meltdown, chen2010side,kocher2020spectre, kocher1996timing, ashokkumar2016highly, irazoqui2014wait,yarom2014flush+, inci2016cache, brumley2005remote, lerman2014power, timon2019non, das2019x, golder2019practical, kovacs2015security, ferrigno2008aes, angel2020private} are just a short list of side-channel attacks that demonstrate how different side effects can be used to lunch fatal attacks.

In 2004, R. Agrawal \cite{asonov2004keyboard} could successfully lead an effort to use acoustic emanations of keystrokes to partially extract the data typed by mechanical keyboards and ATM pads. Since then, tens of published works have been done by researchers to improve the performance of prior works or find a better way for specific scenarios. With the advance of technology, emerging better recording devices, and new ways to record acoustic emanations, acoustic side-channel attacks on keyboards are more feasible and accurate nowadays.

This paper results from our comprehensive literature review covering more than 170 papers. In short, we categorize attack types, briefly explain the techniques used to lunch attacks, compare the success rate of each attack, and discuss the efficiency and applicability of attacks in different scenarios. This work gave us the insight to initiate two projects in this area, and we try to help researchers to understand the area faster and more conveniently. 
\section{Why it is feasible}
Acoustic side-channel attacks on keyboards result from sound and vibration produced by pressing and releasing the keys of keyboards.

\subsection{Distinguishable Sounds}
You have probably noticed the difference between the noise produced by pressing specific keys like Enter and Space. Each keystroke may create some specific sounds for two main reasons: the physics of keyboards and typing styles of people. Physically there is a key plate under the keys of each keyboard. Like when a person plays drum when a person presses a key, it strikes a specific point on the key plate that generates a unique noise\cite{asonov2004keyboard,ponnam2013keyboard}. Other reasons, like microscopic differences in keys' production and the different environment of each key, are also mentioned for the diversity of keystroke noises\cite{asonov2004keyboard}. However, based on our experiments, the differences made by them are not significant enough to launch a successful attack. Type style is the other important factor. People may press different keys with different fingers, pressure, or angles, which in turn, different noises with different intensities might be generated \cite{ponnam2013keyboard}. In addition, people may type some words or syllables with certain styles that some AI techniques can exploit to lunch successful attacks on specific victims. More on this and examples are explained in the following sections.   

\subsection{Practically Recordable Sounds}

There are several scenarios in that attackers can record keystroke Sounds. Here we categorize them into the followings:

\begin{enumerate}
\item \textbf{Physical Proximity}: Today, mobile phones have powerful microphones that have become a good option for recording high-quality stereo sound. Some smartphones have two or three powerful microphones that are good enough to eavesdrop on keystrokes up to 1 meter \cite{zhu2014context} \cite{liu2015snooping} \cite{wang2016accurate} \cite{anand2015bad} \cite{liu2019keystroke} \cite{qin2019lol} \cite{bai2021know}.

In some papers, authors assume that the attacker is physically near the victim and, using a small microphone or a cellphone records the victim's keystroke noises. In \cite{zhu2014context} victims work in public areas like coffee shops or libraries. In \cite{neale2006investigating}, \cite{anand2015bad}, \cite{wang2016accurate} the attacker or his allies can visit/intrude on the victim's office, or leave a smartphone, a hidden microphone, or mp3 players near the victim\cite{qin2019lol}. Some authors consider a scenario where the victim and the attacker are waiting to get a service from a machine by entering a code or password on a keypad\cite{asonov2004keyboard} \cite{liu2019keystroke} \cite{panda2020behavioral} \cite{foo2010timing} \cite{ranade2009acoustic} \cite{cardaioli2019your} \cite{de2019differential} \cite{de2015identification}. ATM machines, POS devices, and digital door locks are examples of such machines.
\begin{figure} [h]
    \centering
    \subfigure[Spying microphone]{\includegraphics[width=0.2\linewidth]{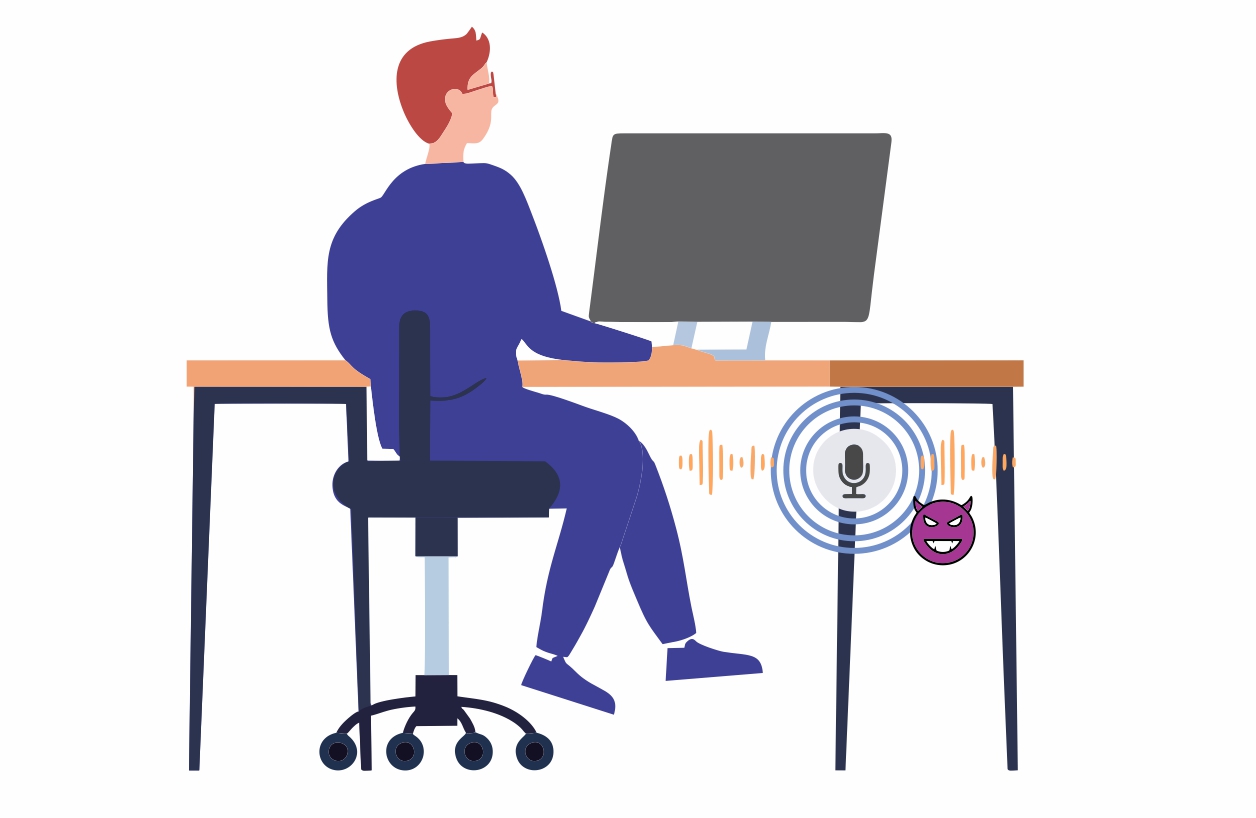}}
    \hfill
     \subfigure[Spying Phone]{\includegraphics[width=0.2\linewidth]{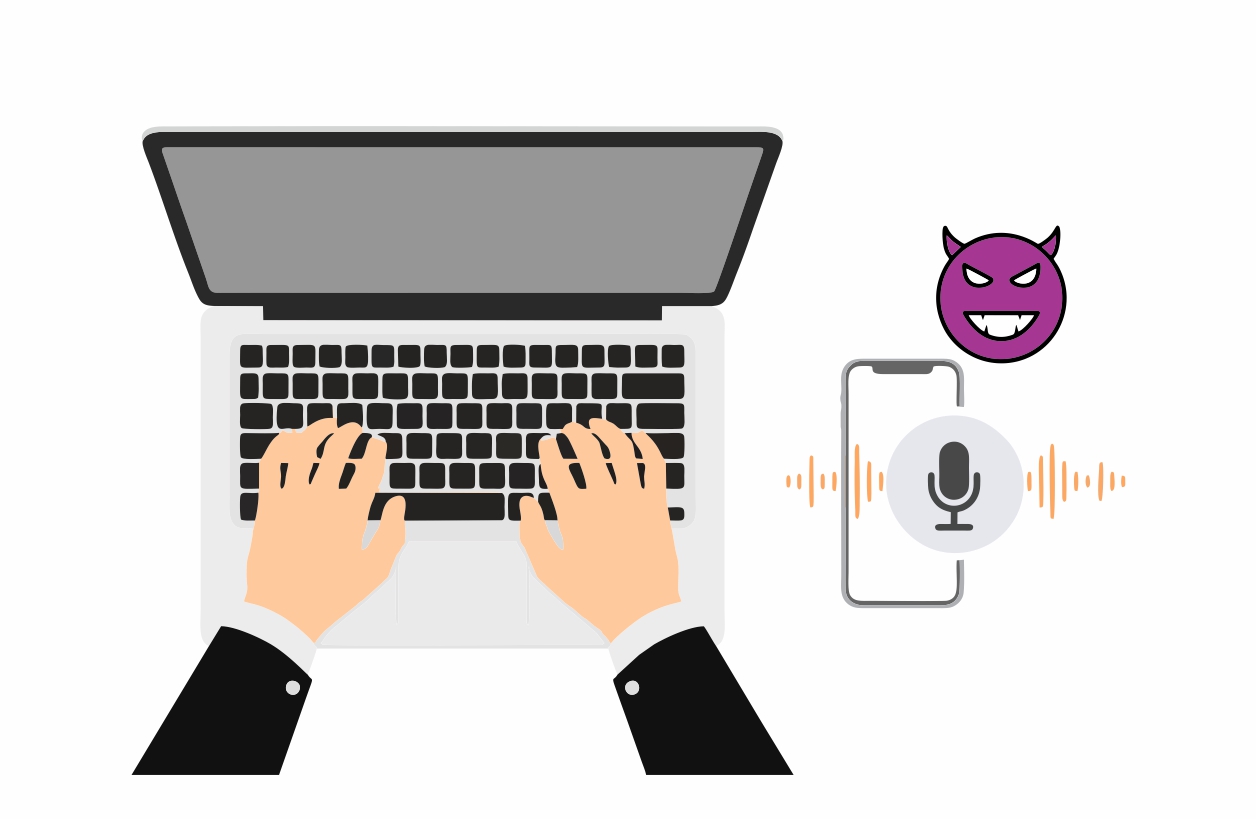}}
    \hfill
    \subfigure[Public Area like a library]{\includegraphics[width=0.2\linewidth]{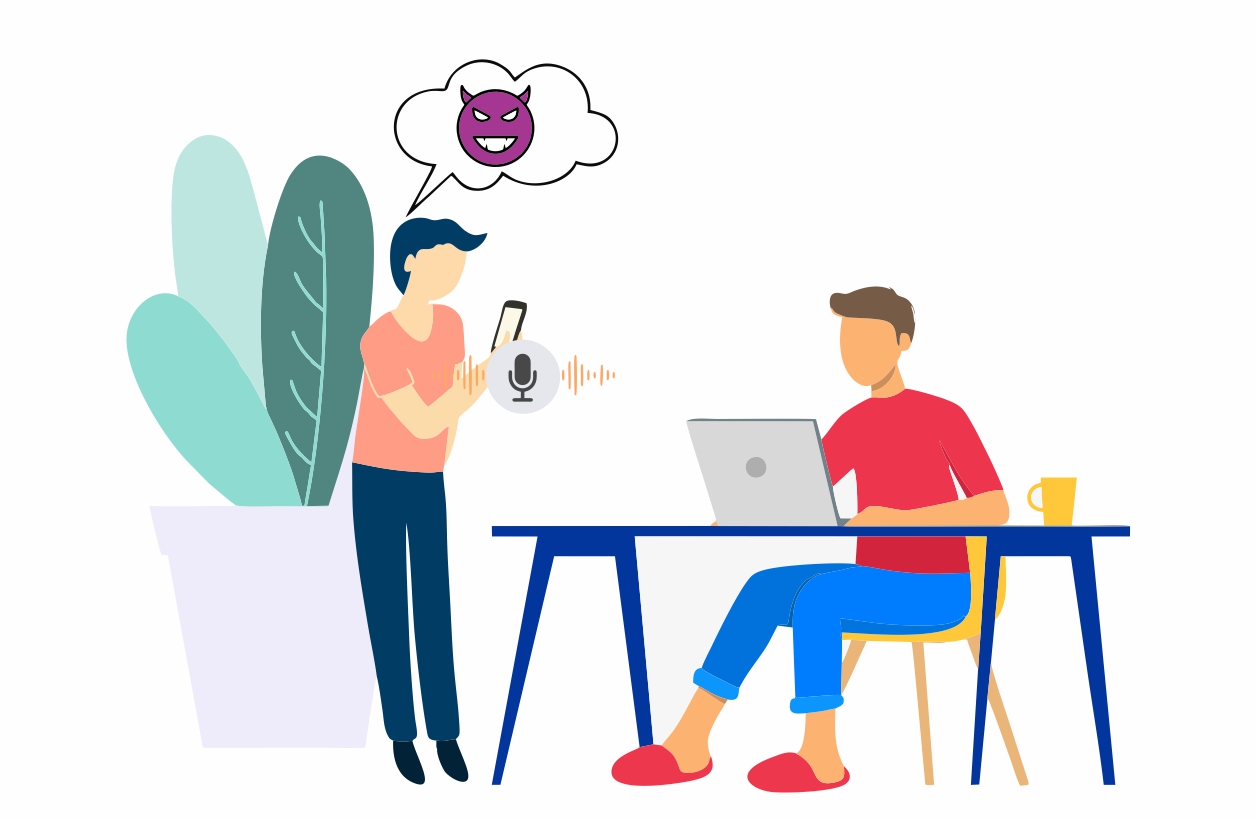}}
    \hfill
    \subfigure[In a queue like a bank ATM]{\includegraphics[width=0.2\linewidth]{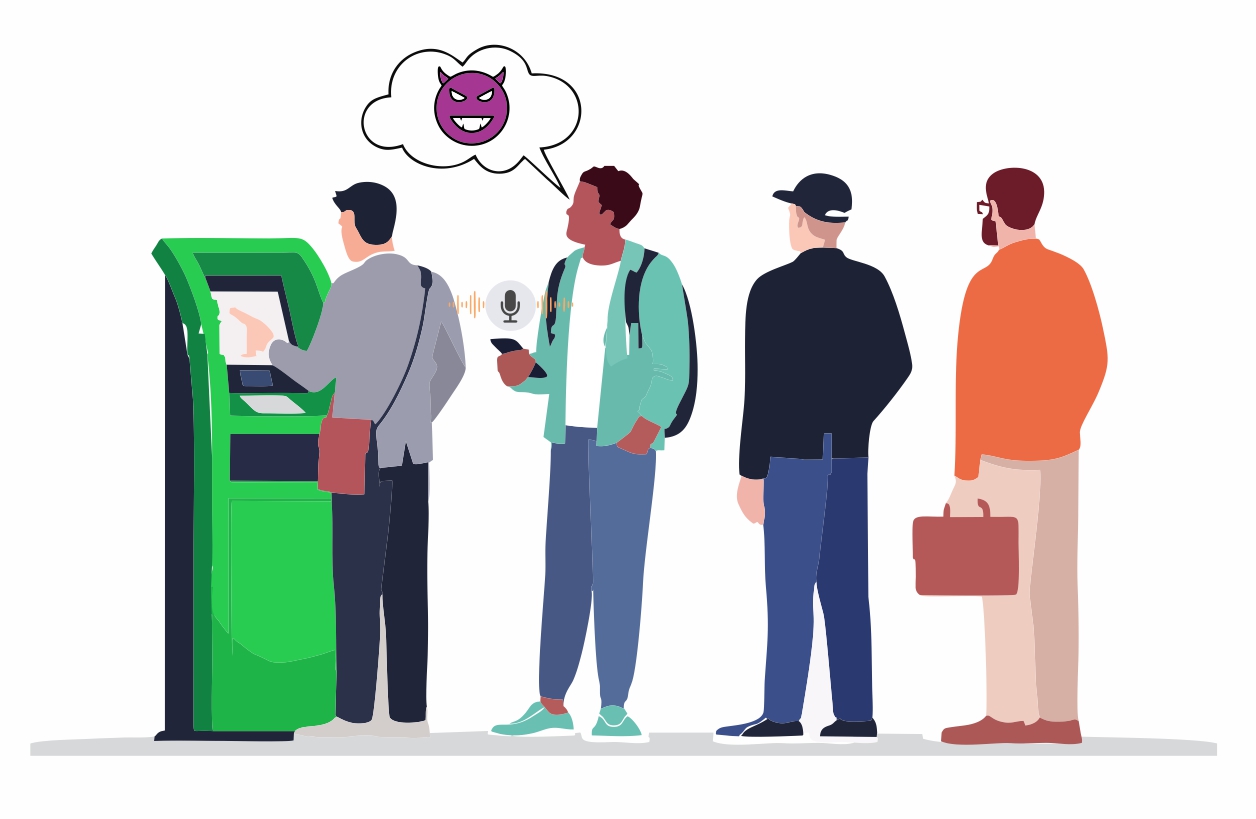}}
    \caption{Examples of Physical Proximity scenarios}
    \label{fig:three-images}
\end{figure}

\item \textbf{Physical access}: in some cases, an attacker may have physical access to the victim's device for a limited time. For example, the victim may have a shared office with the attacker, or the device itself might be a shared device like the computers of public sites in hotels, car dealers, hospitals, banks, and governmental offices. Some special keyboards like pin pads are shared devices by nature. This scenario works for all attack methods which only focus on the physical characteristics of audio signals and keyboards\cite{halevi2012closer} \cite{halevi2015keyboard}.

\begin{figure}[h]
  \centering
  \includegraphics[width=0.4\textwidth]{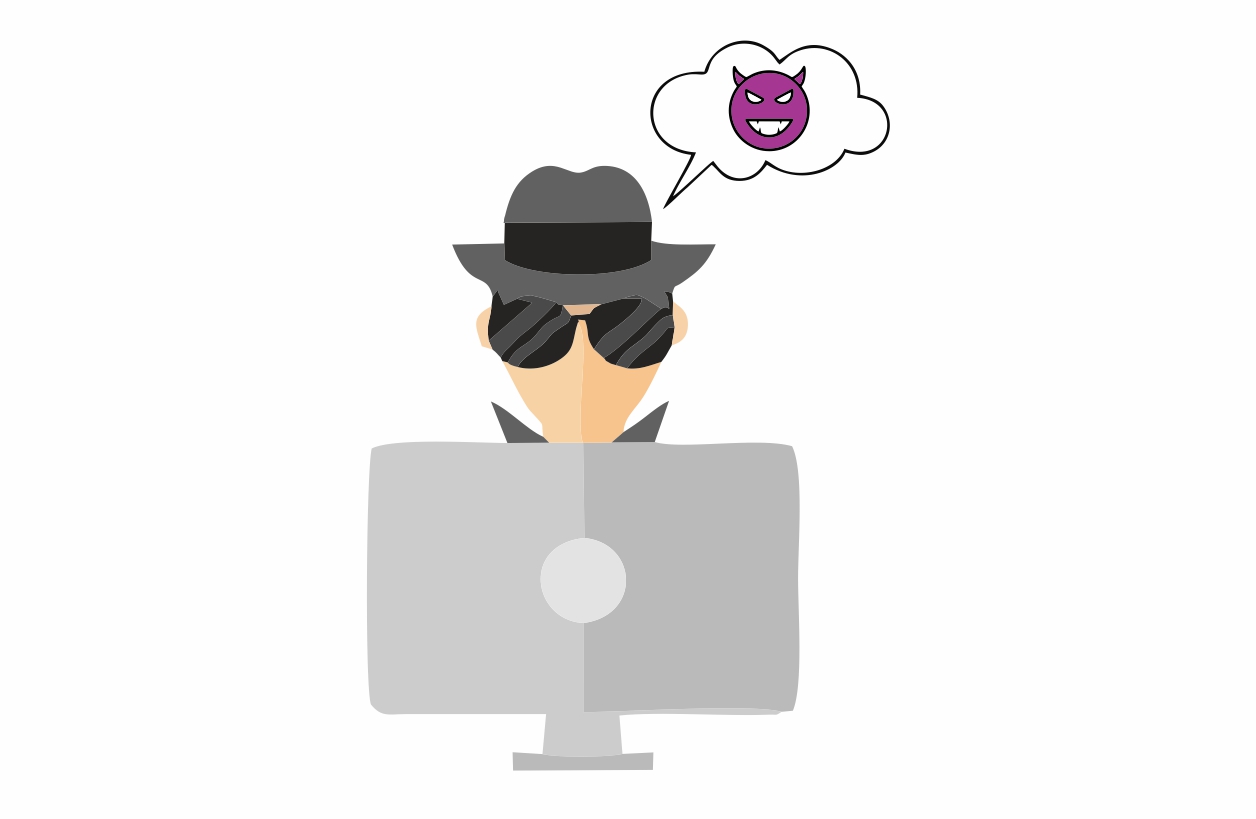}
  \caption{Direct access to the victim's device for a limited time}
  \label{fig:DirectAccess}
\end{figure}

\item \textbf{Indirect physical access}:
The attacker may not gain access to the victim's device, but he can provide a similar device. Although the success rate of attacks may reduce with this approach, it is still a practical attack scenario. Note that most companies and government offices use a limited set of brands and models. It is not hard to find out the common keyboard models used in a company and buy a similar keyboard and lunch experiments with them. 

\item \textbf{Remote indirect access}:
In this method, the attacker tries to record keystroke noises using a medium tool:

\begin{itemize}
    \item Trojans: to accomplish the attack, the attacker may create a malicious application that masquerades as a legitimate app, such as a game, virtual journal, or weather app, which is then uploaded to a typical app marketplace like Google Play and App Store \cite{cheng2020sonarsnoop} \cite{cheng2020acoustic}. One can be a malicious app installed on the victim's phone, \cite{gupta2016deciphering} \cite{zhou2018patternlistener} \cite{giallanza2019keyboard} smartwatch \cite{fang2018no}, or computer. The adversary has different methods of installing this malicious app. Firstly, the victim may naively install the app\cite{simon2013pin} \cite{shumailov2019hearing}. Secondly, social engineering techniques may trick the victim into installing the app. Lastly, the adversary may access the victim's phone somewhere and download the app. Once the malicious application is installed, it requests access to microphone privileges, which users often overlook and grant to the app without question \cite{cheng2020sonarsnoop} \cite{zarandy2020hey} \cite{das2014you}  \cite{anand2018keyboard}. This allows the adversary to use the microphone to record audio without the user's knowledge or consent. Several studies have shown that many popular apps request microphone access and users tend to grant this access without considering the consequences.
    \cite{kelly2010cracking} \cite{wang2016accurate} \cite{schlegel2011soundcomber} \cite{simon2013pin} \cite{slater2019robust} \cite{monaco2018sok} \cite{liu2015snooping} \cite{narain2014single}
    
    \item Hacking third-party applications: Attackers can exploit software bugs to gain unauthorized access to a person's computer or mobile device, allowing them to steal sensitive information or spy on the user. One way attackers can exploit software bugs is by using them to access a device's microphone. Doing so allows them to record ambient sounds without the person's knowledge, including keystroke noises \cite{marques2016snooping}. In \cite{anand2018keyboard}, authors can record victims' ambient sound by using the recorder plugin for Skype named "Supertintin" \cite{anand2018keyboard}.
    
    \item Voice assistant devices: Like any other connected device, voice assistant devices are potentially vulnerable to spying attacks. One potential vulnerability of voice assistant devices is that they can be hacked or compromised through vulnerabilities in their software or firmware, which could allow attackers to remotely access the device, listen to conversations, or perform malicious actions.    
\cite{zarandy2020hey} \cite{cheng2020acoustic}

\begin{figure} [h]
    \centering
    \subfigure[Remote indirect access]{\includegraphics[width=0.3\linewidth]{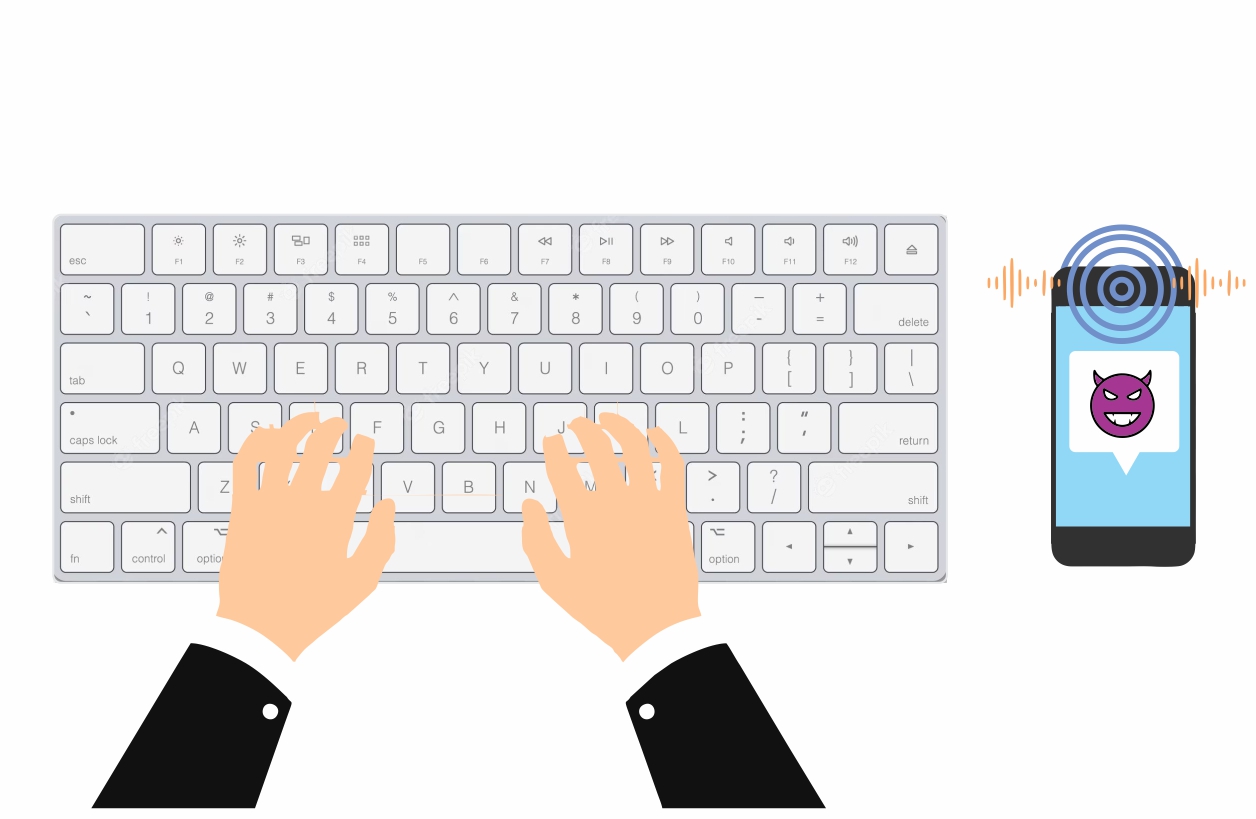}}
    \hfill
     \subfigure[malicious application on Pc]{\includegraphics[width=0.3\linewidth]{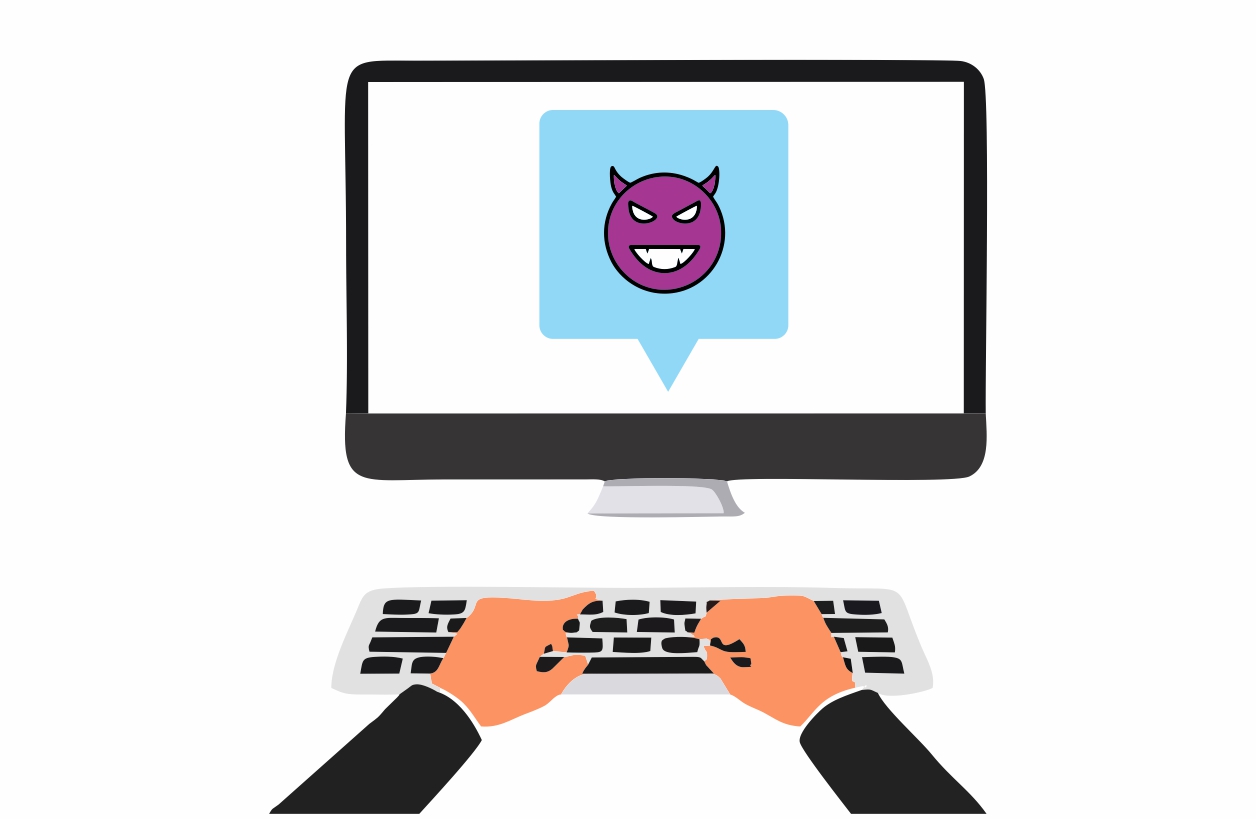}}
    \hfill
     \subfigure[Voice assistant devices]{\includegraphics[width=0.3\linewidth]{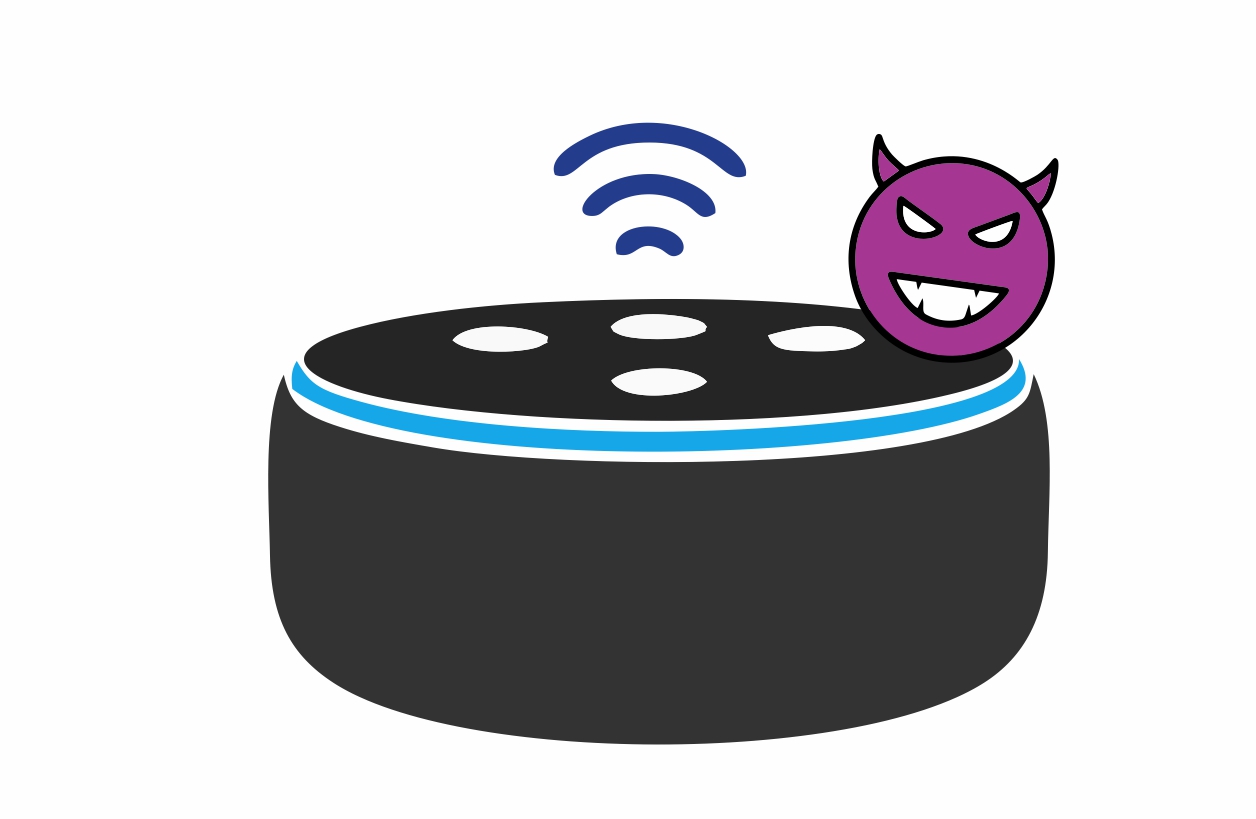}}
     \caption{Examples of Remote indirect access}
    \label{fig: Remoteindirectaccess}
\end{figure}

\end{itemize}

\item \textbf{VoIP and Online meetings}:
Voice over Internet Protocol (VoIP) is a technology that allows voice communication to be transmitted over the Internet instead of traditional telephone lines. Recently, it has become increasingly popular as an alternative to traditional telephone services. VoIP offers a range of additional features and capabilities, such as video conferencing, screen sharing, and instant messaging. Attackers can record audio signals and keystrokes during a Voice-over-IP (VoIP) call or online meetings using software such as Zoom, Skype, or Google Hangouts without having to install malware on the victim's phone\cite{cecconello2019skype}  \cite{compagno2017don} \cite{anand2018keyboard}\cite{sabra2020zoom}. This method differs from other methods in that the attacker can, even during a seemingly legitimate and ordinary online meeting or voice call, record the ambient noise of the victim. 

The attack is made possible by assuming that most users multitask during VoIP calls and online meetings and engage in sensitive activities such as sending emails, chatting, taking notes, paying financial bills, and entering PINs and random passwords. When a user presses a key, it produces a sound transmitted to the other side of the call, allowing attackers to capture the audio packets and remotely decode the keystrokes.

\begin{figure} 
    \centering
    \subfigure[Online Meeting]{\includegraphics[width=0.3\linewidth]{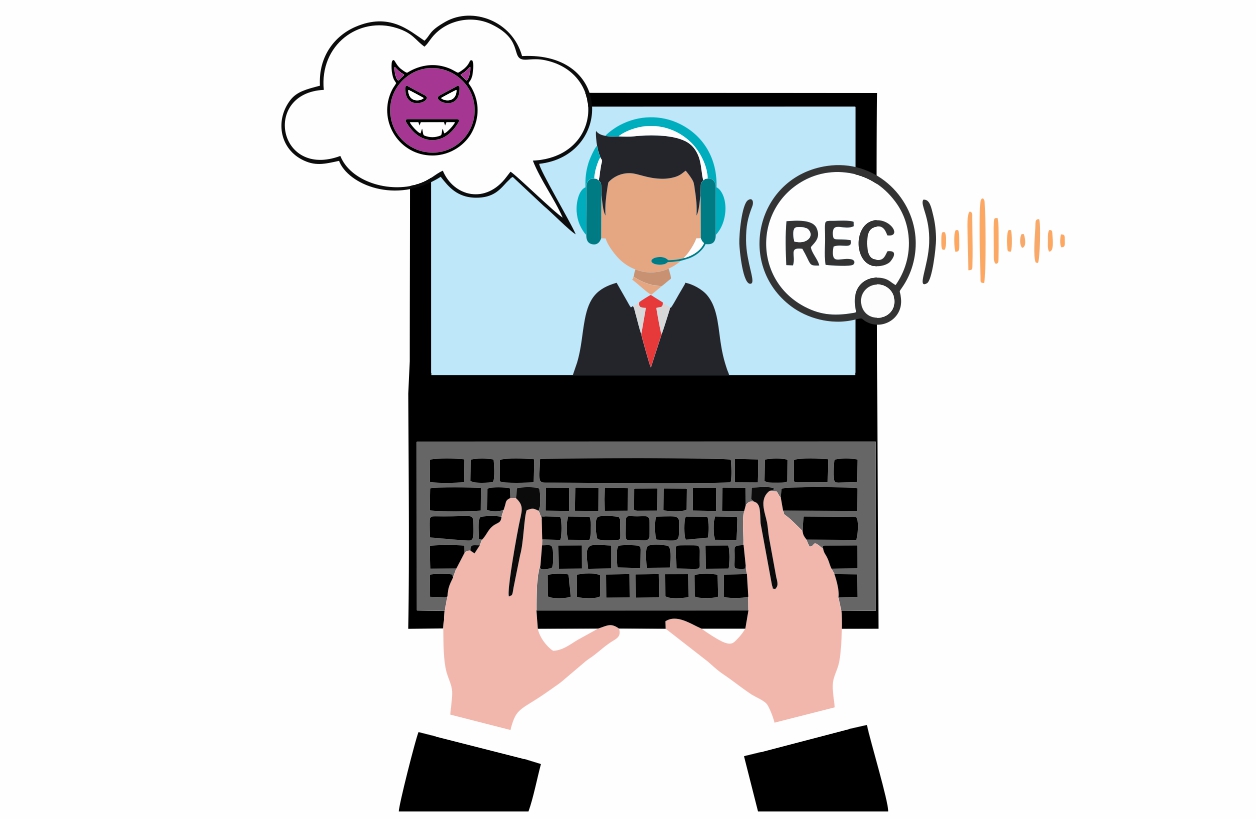}}
    \hspace{1mm}
     \subfigure[VOIP Call]{\includegraphics[width=0.3\linewidth]{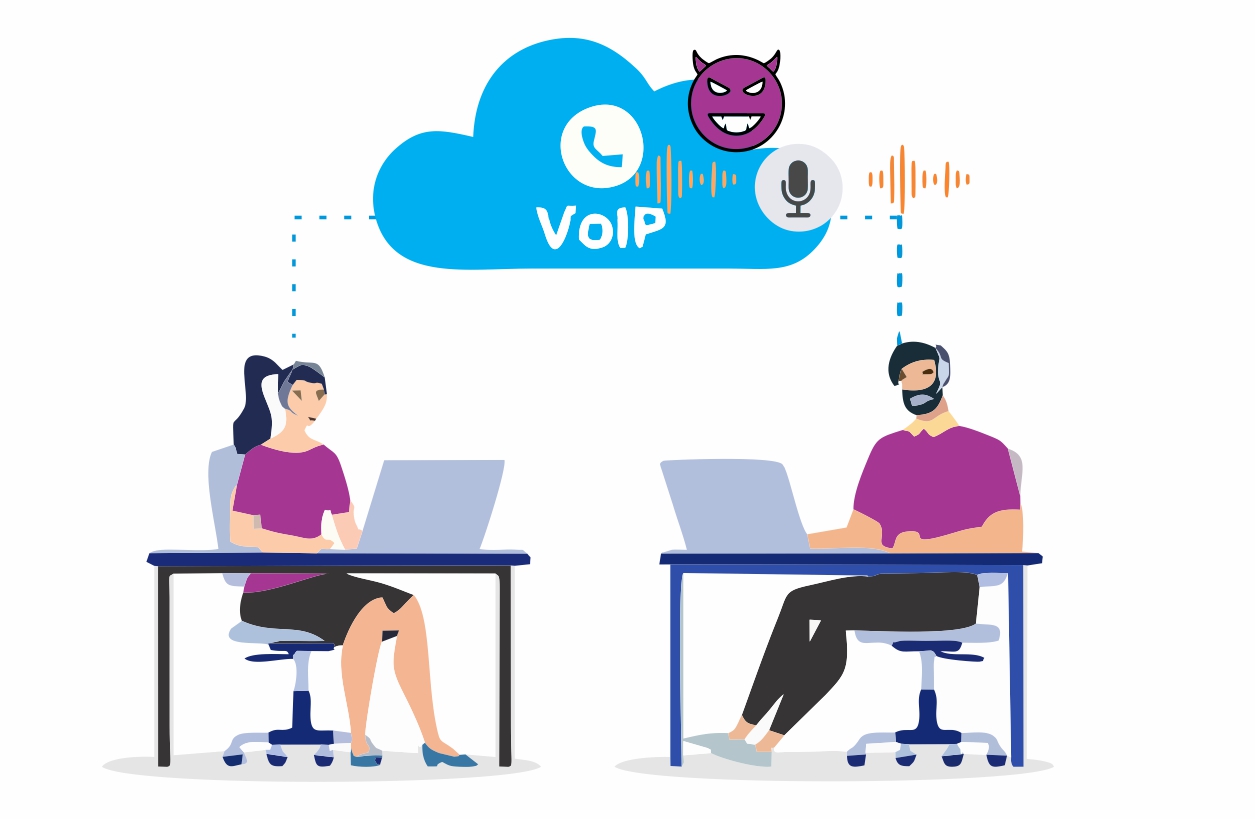}}
        \caption{Example of VoIP and Online meetings}
    \label{fig:VoIPandOnlinemeetings}
\end{figure}
\end{enumerate} 
\section{Attack Surface}
 \subsection{Recording Medium}
 Data collection is the most critical step of an acoustic side-channel attack. Without appropriate data, launching a successful attack is less feasible. In our case, the appropriate data is the clear keystroke sounds; Any other voice is considered noise. Less noisy environments like public libraries or private offices increase the chance of successful acoustic side-channel attacks. Researchers considered a variety of recording devices in the data-gathering stage:
\begin{enumerate}
    \item \textbf{Off-the-shelf microphones}, \textbf{attacker's cellphone} or \textbf{voice recorder}: These devices can be used when the attacker has physical access to the victim's office or when the victim works in quiet public environments like hotel lobbies or libraries. These recording devices should be placed near the victim's keyboard (less than a meter)\cite{zhu2014context} \cite{liu2015snooping} \cite{wang2016accurate} \cite{lu2019keylistener}  \cite{liu2019keystroke} \cite{qin2019lol} \cite{cardaioli2019your} \cite{de2019differential}.
    
    \item \textbf{Victim's cellphone or computer}: Generally, these devices are the better recording devices to gather the required data, because they are the nearest thing to the victim's keyboard. People put their cell phones in close proximity to their keyboards. A laptop's internal microphone is another suitable recording device if the target is a laptop. The downside of this approach is that the attacker must compromise the victim's cellphone or laptop. But, it makes remote attacks possible and usually gathers less noisy data\cite{martinasek2015acoustic} \cite{schlegel2011soundcomber} \cite{foo2010timing}  \cite{bai2021know} \cite{slater2019robust}. It should be mentioned that some researchers assume attackers capture the victim's keystroke sounds during a legitimate-looking online meeting. In this case, even though the attacker uses the victim's computer to gather the keystroke sounds, there is no need to compromise the victim's computer\cite{cecconello2019skype} \cite{compagno2017don}.       

    \item \textbf{Hyperbolic} or \textbf{parabolic microphones}: These are special recording devices that capture weak sound emitting from a far source using a reflective dish (See Figure \ref{fig:ParabolicMic}). The dish focuses incoming sound waves onto a small microphone to capture distant sounds with enhanced clarity. Hyperbolic microphones are commonly used in wildlife recording, surveillance, sports broadcasting, and long-range audio capture applications. They excel at picking up sounds from specific directions while reducing ambient noise and interference. In \cite{asonov2004keyboard}, the authors could launch an acoustic side-channel attack on a keyboard from a 15-meter distance.

      \begin{figure}
        \centering
        \includegraphics[width=0.25\linewidth]{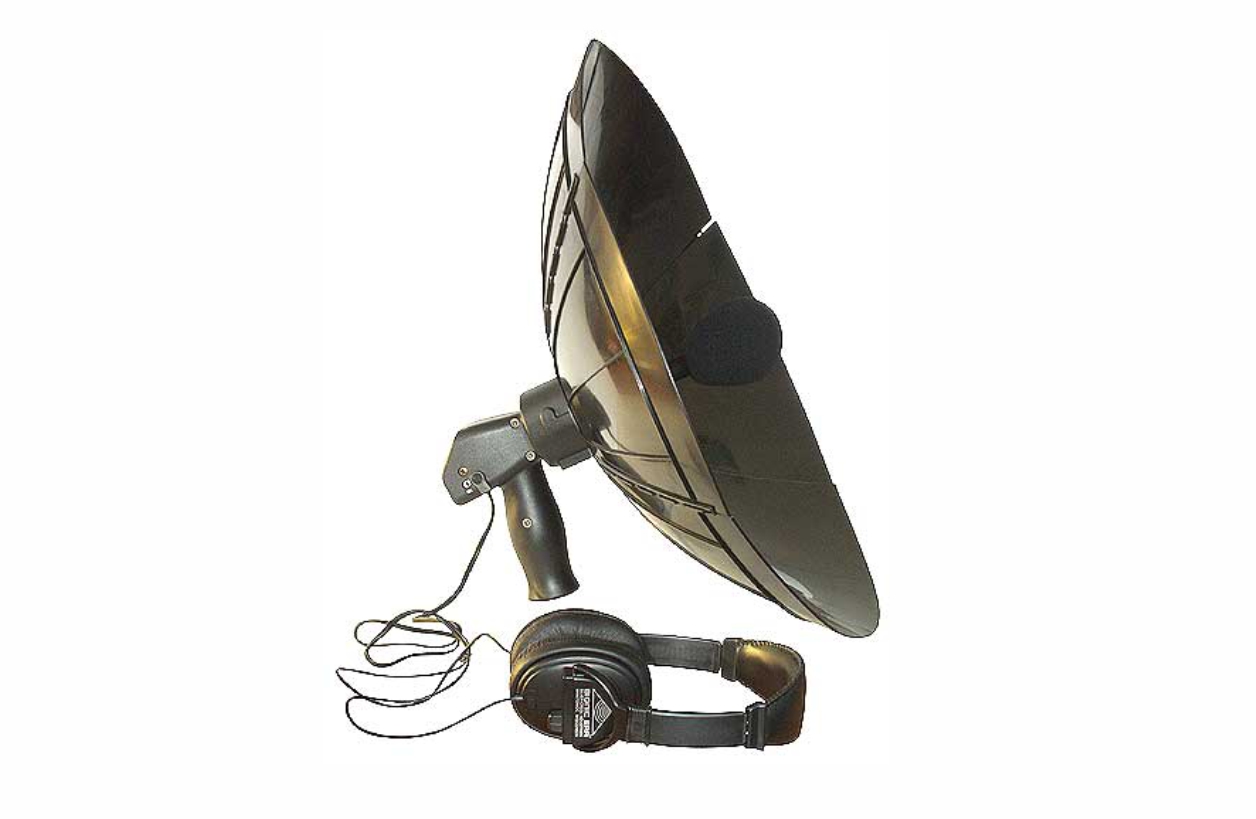}
       \caption{Parabolic Microphone}
    \label{fig:ParabolicMic}
      \end{figure}
      
\end{enumerate}

 \subsection{Keyboard Types}
\begin{enumerate}

\item \textbf{Mechanical PC keyboards}: 
Using a commodity mechanical keyboard (See Figure \ref{fig:MeckKeyboards}) to interact with PCs, Laptops, and Notebooks is common. This type of keyboard produces more audible noises, making them more vulnerable to acoustic side-channel attacks. Hence, most articles like \cite{asonov2004keyboard}  \cite{qin2019lol}
\cite{ponnam2013keyboard} \cite{fiona2006keyboard} \cite{berger2006dictionary} \cite{halevi2012closer} 
\cite{wit2014all} \cite{halevi2015keyboard} \cite{kelly2010cracking} \cite{rosmansyah2017microphone} \cite{zhu2014context}
\cite{wang2016accurate} have focused on them. In general, laptop keyboards have softer keys and emanate fewer noises. \cite{bai2021know} \cite{liu2019keystroke} Therefore, they are less vulnerable to many attacks, and most researchers have reported lower success rates than PCs' mechanical keyboards \cite{halevi2012closer} \cite{halevi2015keyboard} \cite{asonov2004keyboard}

\begin{figure}[h]
  \centering
  \includegraphics[width=0.6\textwidth]{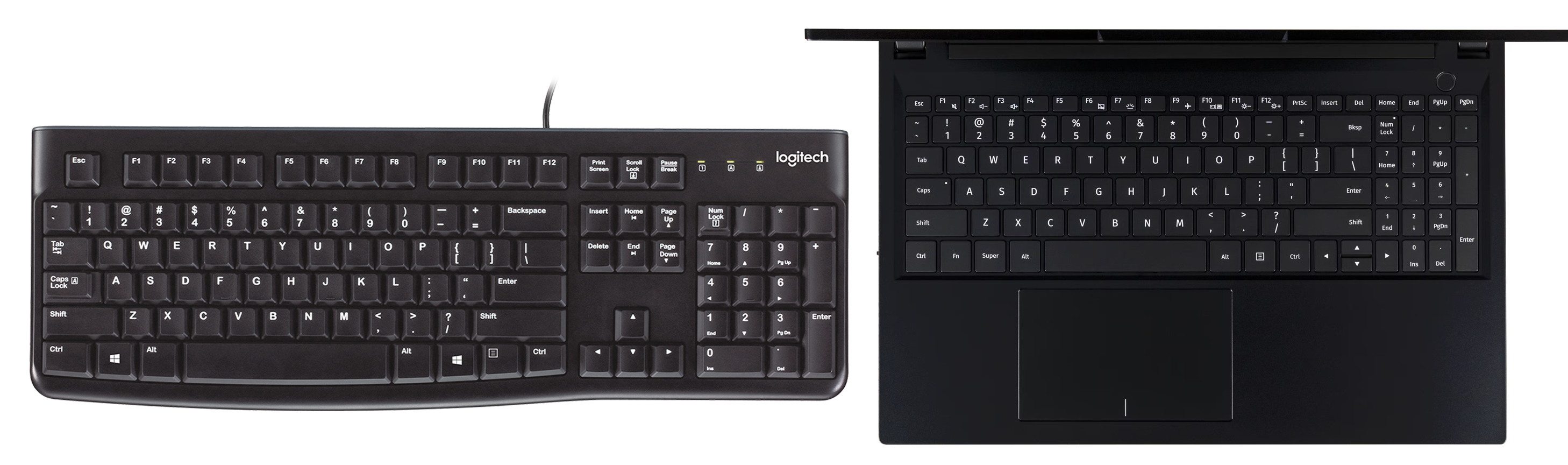}
  \caption{Mechanical keyboards}
  \label{fig:MeckKeyboards}
\end{figure}

\item \textbf{Virtual Keyboards}:
Most modern smartphones use software-based virtual Keyboards, which can be used to enter PIN codes, authentication patterns, or text (See Figure \ref{fig:VirKey}). This type of keyboard is also vulnerable to acoustic side-channel attacks. In \cite{gupta2016deciphering}  \cite{shumailov2019hearing}, the authors target the virtual keyboard of Android smartphones or tablets. 
Zarandy et al. implemented an attack on the virtual keyboard of an Android smartphone, which extracts the Pin codes\cite{zarandy2020hey}.
In another work\cite{cheng2020sonarsnoop}, the unlock pattern of Android phones was attacked. 
In \cite{schlegel2011soundcomber}, the authors designed a malicious app (like a Trojan) that legitimately grants mic access from victims, records phone calls, and extracts sensitive information like credit card numbers when victims need to enter them over a phone call. A series of works \cite{lu2019keylistener}, \cite{kim2020tapsnoop}, \cite{yu2019indirect}, \cite{narain2014single} tried to extract keystrokes on touchscreen QWERTY keyboards. 
\begin{figure}[h]
  \centering
  \includegraphics[width=0.4\textwidth]{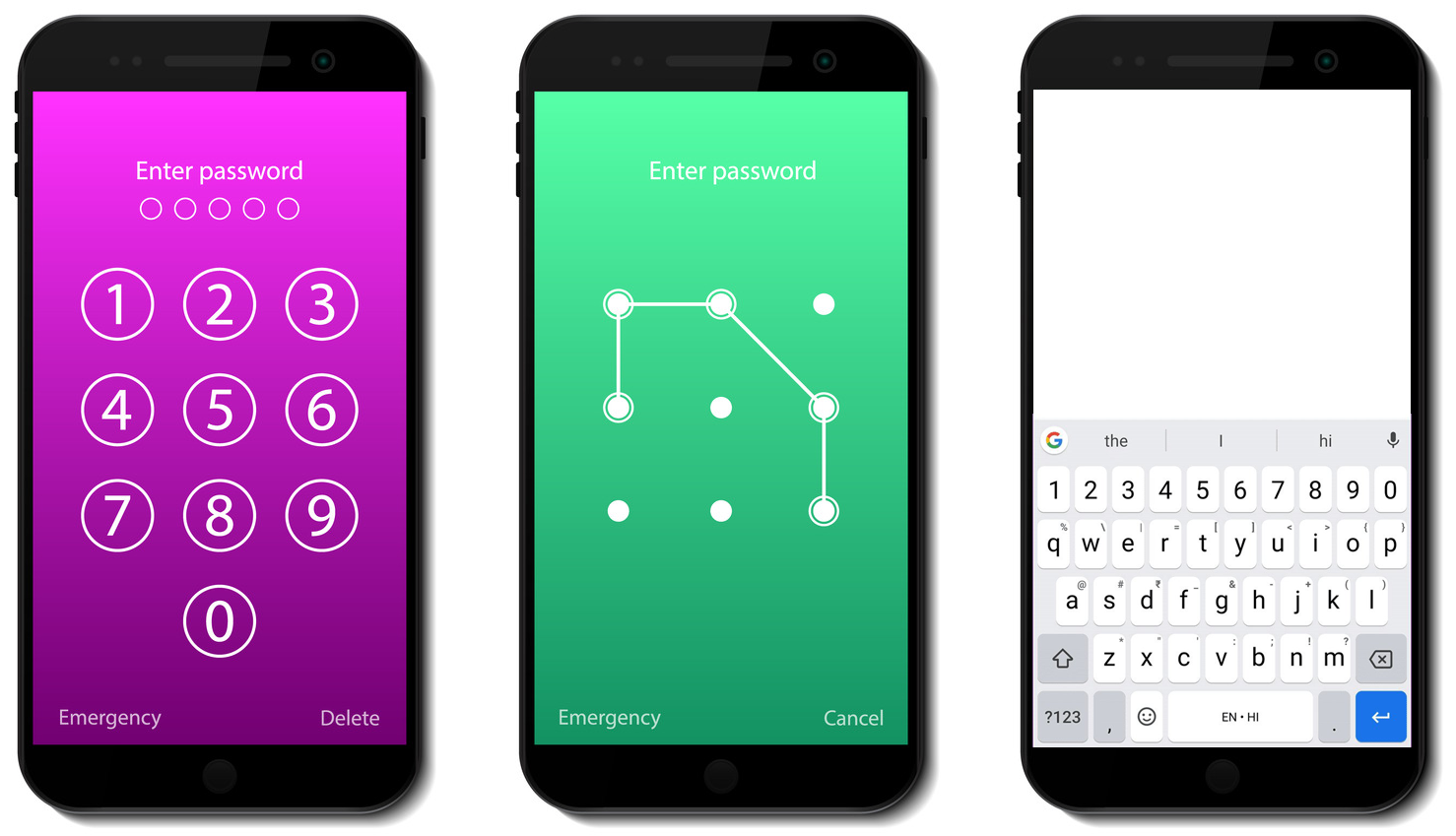}
  \caption{Virtual keyboards}
  \label{fig:VirKey}
\end{figure}

\item \textbf{PIN Keypads}:
PIN keypads (See Figure \ref{fig:PEDdevices}) are electronic devices used for entering personal identification numbers (PINs). They are commonly found in a wide range of security systems and payment processing applications, such as automated teller machines (ATMs), point-of-sale (POS) terminals, and door access control systems. PIN keypads typically consist of a panel of a few buttons, and people usually use only one figure to enter sensitive information one by one. Consequently, they are more attractive and vulnerable to acoustic side-channel attacks \cite{panda2020behavioral}, \cite{foo2010timing}, \cite{cardaioli2019your}, \cite{ranade2009acoustic}.

\end{enumerate} 
\subsection{Target Data}
Researchers have tried to increase their attack success rate by limiting the target of the attacks. Here is the list of some data types that were the target of attacks:
\begin{enumerate}
    \item \textbf{Unlock patterns}: A phone unlock pattern is a security feature commonly found on Android devices. It involves tracing a specific pattern on a grid of dots displayed on the phone's lock screen. This pattern serves as a passcode to unlock the device and access its contents. The papers \cite{cheng2020sonarsnoop} and \cite{zhou2018patternlistener} are two examples of papers that focused on Android unlocking patterns.  
    \item \textbf{Single keys}: Many beginners tend to press keyboard keys individually, allowing attackers to distinguish each key's sounds clearly. Even professional typists may create a noticeable gap between keystrokes when entering one-time codes or random passwords, which enables the same thing. Assuming there are some scenarios that allow us to gather clear sound of keystrokes, several works have been focused on detecting single keys \cite{fiona2006keyboard}\cite{narain2014single}\cite{asonov2004keyboard} \cite{liu2015snooping} \cite{ellispredicting} \cite{rosmansyah2017microphone}\cite{martinasek2015acoustic}. 
    \item \textbf{passwords}: 
    Passwords, the common login credentials, are the most attractive data for most attackers. Hackers target login credentials to gain unauthorized access to various accounts, such as email, social media, banking, or corporate systems. Hence it has been the subject of many research \cite{zhu2014context} \cite{kelly2010cracking} \cite{zhuang2009keyboard} \cite{berger2006dictionary} \cite{bai2021know}  \cite{slater2019robust} \cite{wit2014all}. Some researchers have focused on special types of passwords. For example, Panda et al. targeted only numerical passwords\cite{panda2020behavioral}; Halevi et al. worked on passwords consisting of lower-case English alphabets \cite{halevi2012closer}; and Ponnam et al. investigated extracting passwords composed of English words\cite{ponnam2013keyboard}. 
    \item \textbf{Text}: 
    Texts may also contain sensitive or private information. Consequently, some researchers have been focused on extracting typed text from ambient noises\cite{zhu2014context} \cite{kelly2010cracking} \cite{zhuang2009keyboard} \cite{berger2006dictionary} \cite{bai2021know}  \cite{slater2019robust} \cite{wit2014all} \cite{zhuang2009keyboard} \cite{kelly2010cracking} \cite{gupta2016deciphering}. Typically excluding typos, a text consists of punctuation marks and words that can be found in a dictionary. Moreover, texts adhere to a structure and often follow some recognizable patterns. These features allow researchers to improve their method by filtering out less likely options. 

    \item \textbf{PIN Code}:
    PIN stands for Personal Identification Number. PIN codes are a particular type of password consisting of some digits; usually, the length of PIN codes in a system is fixed and known, and they are used as a convenient way for authenticating people. PIN codes are widely used in ATM and POS devices to authorize transactions and access funds,  mobile devices as unlocking codes, and security Systems and digital locks to grant access to restricted areas or buildings. Hence, we categorized them as an independent category. Researches like \cite{simon2013pin} \cite{schlegel2011soundcomber} \cite{anand2018keyboard} \cite{panda2020behavioral} \cite{shumailov2019hearing} focused on PIN codes.

    \item \textbf{Miscellaneous}:
    In 2015, Anand et al.\cite{anand2015bad} designed an acoustic side-channel attack against the Tapping-based Rhythmic Password mechanism\cite{wobbrock2009tapsongs}, which Wobbrock had introduced in 2009.
    Wang et al. proposed a method to counter fast typists in 2016. When a typist presses two consecutive keys in a short interval, the signals of the two keystrokes become mixed. As a result, methods that rely on individual keystroke signals are ineffective. Their approach works by considering two combined keystrokes and utilizes the blind signal separation technique to separate the mixed signals\cite{wang2016accurate}.
\end{enumerate}
\section{Attack Strategies}
This section explores the predominant approaches for identifying keystrokes using acoustic signals. These procedures may involve several stages, including the keystroke signal detection stage. Note that the accuracy of the keystroke signal detection stage significantly influences the overall outcome \cite{zhuang2009keyboard}. We classify the attack strategies into the following categories.

\begin{figure}[h]
  \centering
  \includegraphics[width=0.6\textwidth]{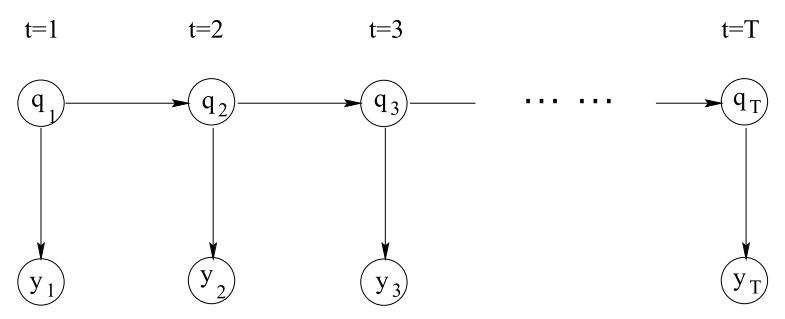}
  \caption{Illustration of a Hidden Markov Model (HMM) trace, with each vertical segment corresponding to a time step. Within each time step, the upper node q\textsubscript{t} represents a character pair, while the lower node y\textsubscript{t} represents the observable variable indicating the latency between consecutive keystrokes \cite{song2001timing}.}
  \label{fig:hmms}
\end{figure}

\subsection{Timing-based Analysis}
Timing information is a type of side-channel information related to typing patterns. Timing information includes the time between the start or end of two keystrokes, the duration of each key press, the time from keystroke push to keystroke release, and so on \cite{foo2010timing, liu2019keystroke}.
As illustrated in Figure \ref{fig:hmms}, Foo et al. \cite{foo2010timing} applied the techniques developed by Song et al. \cite{song2001timing}, utilizing a Hidden Markov Model (HMM) for analyzing inter-keystroke timings in the context of inter-keystroke timing attacks.
\cite{liu2019human} suggests user-independent inter-keystroke timing attacks on PINs using a human cognitive model. The findings indicate that their attack methods yielded favorable performance results.

\subsection{Frequency-based Analysis}
This technique focuses on analyzing acoustic signals in the frequency domain. Typically, researchers employ the Fast Fourier transform (FFT) to convert signals from the time domain into the frequency domain, which serves as the base of their recognition system. Additionally, researchers leverage cepstrum features, a common choice in speech analysis and recognition systems. Empirical evidence demonstrates that cepstrum features, especially mel-frequency cepstral coefficients (MFCCs) \cite{zhuang2009keyboard}, tend to outperform plain FFT coefficients when processing sound signals. MFCC applies the Fourier transform to yield the frequency content of the input signal. The attackers require these features from the signal data to detect and classify each keystroke.

\subsubsection{Signal Processing and Statistics Techniques}
In \cite{panda2020behavioral}, Panda et al. propose a side-channel attack on a 4–6 digit random PIN key and achieve a 60\% accuracy in recovering them. Their sample victims were asked to memorize a six-digit number and type it repeatedly to adapt themselves to this new PIN sequence. Then, they captured the victims' keystroke sounds to build their dataset. They generated a feature vector of recorded signals using a Fast Fourier Transform (FFT) and used it as input for various techniques, such as Cross-Correlation and Euclidean distance.
The study by Halevi et al. \cite{halevi2015keyboard} introduces a time–frequency decoding technique for identifying passwords. It also meticulously investigates the influence of typing style on detection accuracy. The findings reveal that when employing the same typing style (hunt and peck) for both data training and decoding, the optimal success rate for accurately detecting a typed key stands at 64\% per character. Furthermore, the study demonstrates that altering the typing style to touch typing during the decoding phase diminishes the success rate to approximately 40\% per character. This approach can potentially decrease the entropy of the search space for random passwords by up to 57\% per character.

\subsubsection{Classical Machine Learning-based Techniques}
Machine learning is crucial in extracting pertinent features from acoustic data, facilitating more accurate key recognition. These models are trained to discern patterns within acoustic signals linked to different keystrokes, enabling them to predict the pressed keys effectively. Machine learning algorithms, including Support Vector Machines (SVM), Random Forests, and neural networks, can be trained using labeled acoustic data to classify recorded sounds into specific key presses. Furthermore, machine learning models can be deployed for real-time keystroke detection and inference, making them valuable tools that raise concerns about potential security breaches.
\cite{zhuang2009keyboard} uses three different ML methods, including Neural Network, Linear classification(Discriminant), and Gaussian Mixture, to classify acoustic signals. 
In the study by Anand et al. \cite{anand2018keyboard}, they utilize FFT coefficients and MFCC as distinguishing attributes for individual keystrokes. These features are employed in supervised machine learning methods, including Logistic Regression and Random Forest, to analyze random passwords and numeric PINs.
In the paper by Halevi et al. \cite{halevi2012closer}, they employ a method based on frequency-domain features. They utilize Mel-Frequency Cepstral Coefficients (MFCC) as inputs for neural networks and apply a 10 ms window with a 2.5 ms window step size, computing 13 MFCCs per window. They analyze a total duration of 40 ms for each keypress event. They use Matlab's \textit{newpnn()} function to create the neural network.
In \cite{asonov2004keyboard}, a proposed attack utilizes a neural network for recognizing the pressed keys. The authors employed the JavaNNS neural network simulator to construct a backpropagation neural network. They utilized the Fast Fourier Transform (FFT) extracted from an 8–10 ms window around the key press peak as the feature. 
Cecconello et al. \cite{cecconello2019skype} employed a Logistic Regression classifier to categorize a dataset containing ten samples for each of the 26 English alphabet keys. This classification was carried out using a 10-fold cross-validation approach. The authors assessed the classifier's accuracy using various spectral features, including FFT coefficients, cepstral coefficients, and MFCC.

\begin{figure} 
    \centering
    \subfigure[The spectrogram of L key acoustic trace.]{\includegraphics[width=0.4\linewidth]{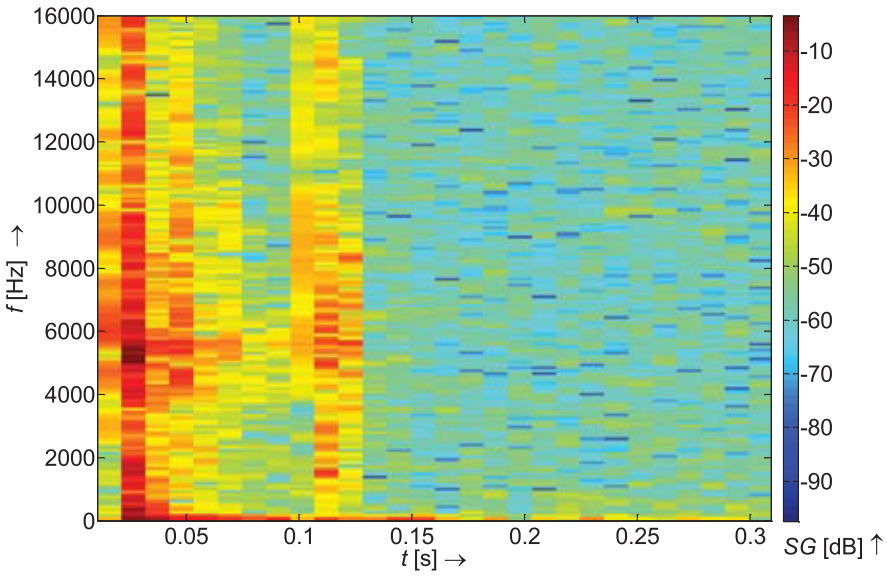}}
    \hspace{1mm}
     \subfigure[The spectrogram of D key acoustic trace.]{\includegraphics[width=0.4\linewidth]{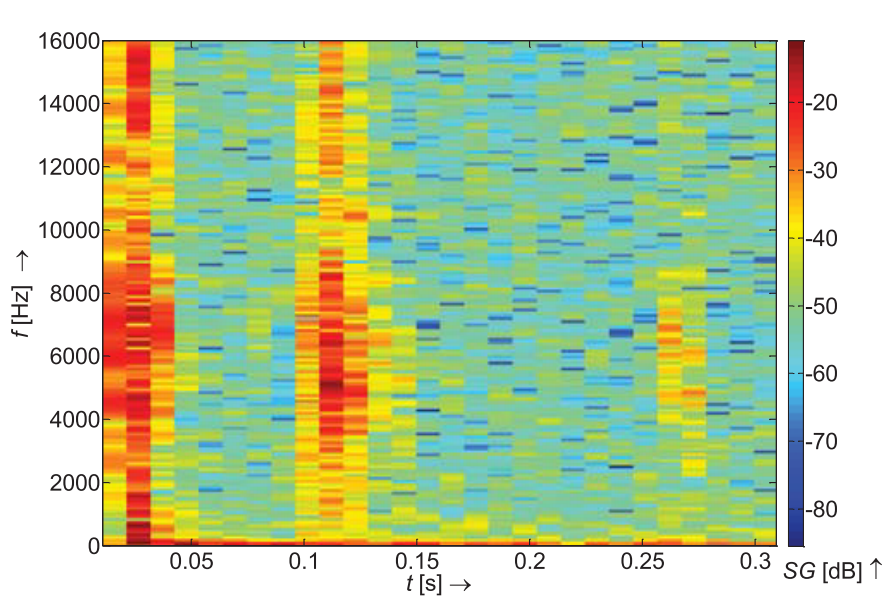}}
        \caption{Spectrograms depicting two distinct keystrokes. \cite{martinasek2015acoustic}.}
    \label{fig:spectrogram}
\end{figure}

\subsubsection{Deep Learning-based Techniques}
\begin{figure}[h]
  \centering
  \includegraphics[width=0.6\textwidth]{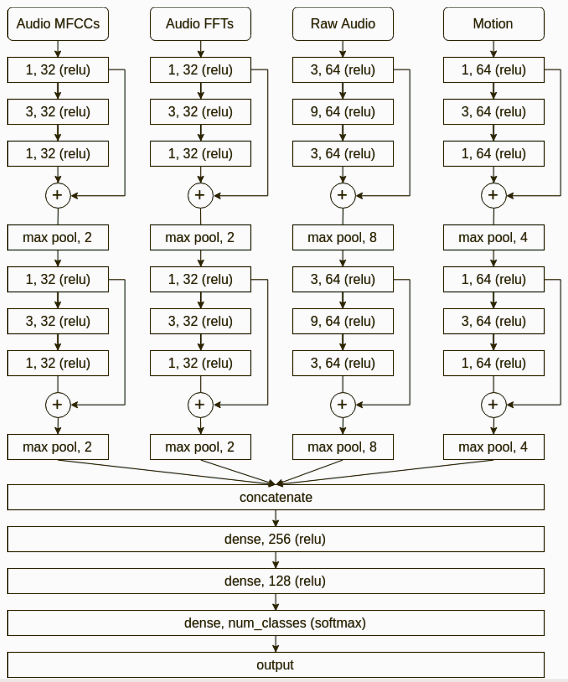}
  \caption{Convolutional network architectures used for keystroke classification \cite{giallanza2019keyboard}.}
  \label{fig:deeplearning}
\end{figure}

Applying FFT to the signal gives us only frequency values, and we lose track of time information. The system won’t be able to tell what was recorded first if we use these frequencies as features. Here, Spectrograms come into the picture. The visual representation of frequencies of a given signal with time is called a Spectrogram. The idea is to break the audio signal into smaller frames(windows) and calculate DFT (or FFT) for each window. This way, we will get frequencies for each window, and the window number will represent the time. In a spectrogram representation plot, one axis represents the time, the second axis represents frequencies, and the colors represent the magnitude (amplitude) of the observed frequency at a particular time. (Bright colors represent strong frequencies).  This is a 2D matrix representing the frequency magnitudes and time for a given signal. Now, think of this spectrogram as an image. This reduces the problem to an image classification problem.
Harrison et al. \cite{harrison2023practical} employed a deep learning model to classify laptop keystrokes. They captured keystroke sounds using the integrated microphone of a smartphone and utilized mel-spectrograms as input for an image classification model named \textit{CoAtNet}. Their proposed classifier achieved an accuracy of 95\% under certain conditions, all without utilizing a language model. Furthermore, they conducted training on keystroke sounds recorded through the video-conferencing software Zoom, attaining an accuracy of 93\%. 
As shown in Figure \ref{fig:deeplearning} in the study by Giallanza et al. \cite{giallanza2019keyboard}, researchers designed a system that combined convolutional and recurrent neural networks for keystroke detection and identification, respectively. This system was tested in an experiment involving 20 participants who typed naturally while conversing. The findings indicated that mobile phone arrays could identify approximately 41.8\% of keystrokes and 27\% of typed words accurately in a noisy setting, even without personalized user training. To assess the potential security implications of this attack, the authors transformed the machine learning models into a real-time system capable of distinguishing keystrokes using an array of mobile phones. The system's performance was evaluated through trials with a single user typing under varying conditions.

In the paper by Akinbi et al. \cite{akinbi4431909password}, audio signals are initially transformed into audio images using the spectrogram representation of the Mel-Frequency Cepstral Coefficients (MFCC) method. Subsequently, a vision transformer-based approach called \textit{ConvMixer} is employed on these spectrogram images to identify passwords based on their audio characteristics. The authors achieved an average accuracy score of 92.44\% in their experimental work. Additionally, a comparison is made with two pre-trained convolutional neural network (CNN) models, namely \textit{ResNet18} and \textit{VGG16}. This comparison underscores the potential of the proposed approach as a compact acoustic keylogger system.
Toreini and colleagues \cite{toreini2015acoustic} harnessed advanced techniques in signal processing and machine learning to discern individual Enigma keys.  Their experimentation involved the detection of four distinct peaks within each sound sample, employing an LPC-based method in the preprocessing stage. In the feature extraction process, they derived MFCC (Mel-frequency cepstrum coefficients) from the audio samples. The highest recognition rate of 92.18\% was achieved by utilizing an Artificial Neural Network classifier. 
In the paper by Martinasek et al. \cite{martinasek2015acoustic}, a spectrogram (as shown in Figure \ref{fig:spectrogram}) serves as the input for a standard two-layer neural network trained using the backpropagation learning algorithm, with no additional tools applied to enhance classification results. The recording setup involved a laptop with an integrated microphone in an office setting. The resulting average success rate for first-order classification in this experiment was 72.3\%.

\subsection{Geometry-based Attacks}
These techniques are based on the geometric positions of keys on keyboards and are employed to estimate the positions of pressed keystrokes \cite{zhu2014context}. A key can be identified by computing the differences in arrival times of a sound wave at two or more microphones \cite{fiona2006keyboard}.
Geometry-based approaches can detect and identify pressed Keys Regardless of the type and meaning of the entered text, so they can work on random texts such as passwords. Using various approaches such as Time Difference of Arrival (TDoA), Triangulation, and differential audio analysis (DAA), attackers estimate the candidate area of a typed key. 

\begin{figure} 
    \centering
    \subfigure[Geometrical TDoA measured by a pair of microphones.]{\includegraphics[width=0.4\linewidth]{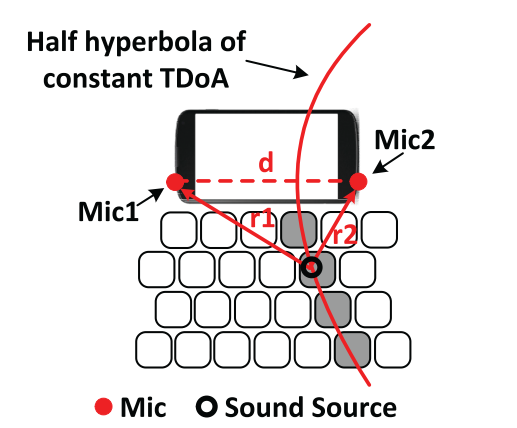}}
    \hspace{1mm}
     \subfigure[Theoretical key groups and corresponding half hyperbolas.]{\includegraphics[width=0.4\linewidth]{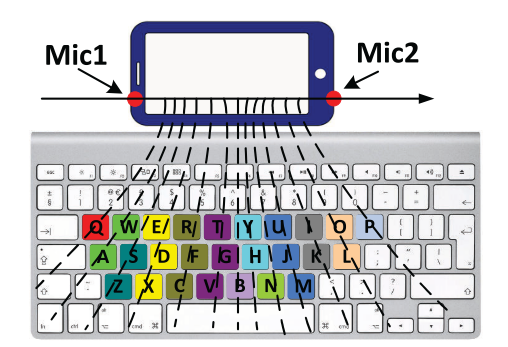}}
        \caption{This is an illustration of the geometrical TDoA on a single phone and the theoretical key groups \cite{liu2015snooping}.}
    \label{fig:tdoa}
\end{figure}

\subsubsection{Time Difference of Arrival and Triangulation Attacks}
Zhu et al. \cite{zhu2014context} introduces an innovative technique that offers context-independent keystroke recovery, even when inputted randomly. Their methodology involves employing two or more smartphones to capture acoustic emanations resulting from keystrokes\cite{rosmansyah2017microphone}. This technique then processes the collected acoustic signals, employing a geometrical framework that leverages the Time Difference of Arrival (TDoA) methodology to pinpoint the physical location of the pressed key. The TDoA technique calculates the variations in the distance between the key and each microphone, ultimately enabling the estimation of a set of potential key positions on a plane. This candidate key set is progressively narrowed down through the collaborative effect of multiple microphone pairs. Their experimental results demonstrate an impressive success rate of over 72.2\% in accurately recovering keystrokes.
Cheng and colleagues \cite{cheng2022dictionary} present an approach for determining keystroke positions through the Time Difference of Arrival (TDoA) technique and subsequently reconstructing words via a dictionary. This attack employs just a single phone and doesn't necessitate any training. The outcomes reveal a 45.2\% success rate in correctly identifying words within the top 50 candidates across the tested vocabulary. Moreover, for words exceeding 10 characters, the success rate improves to 46.3\% within the top 25 candidates.
As shown in Figure \ref{fig:tdoa}, by employing the Time Difference of Arrival (TDoA) technique to locate keystrokes, researchers can eliminate the need for labeled training data or linguistic context. Experiments conducted with three different types of keyboards and off-the-shelf smartphones demonstrate scenarios where this system can recover 94\% of keystrokes, as reported in \cite{liu2015snooping}.

Another related approach to this kind of attack is an acoustic triangulation attack. This attack involves determining an object's position by measuring acoustic waves produced by a keystroke \cite{canistraro1996projectile}. Triangulation is a method for determining the distance of a point through the application of triangle-based principles. The distance is considered as one of the sides of a triangle, computed by measuring its angles and other sides. Triangulation is a widely employed approach for object localization, finding applications in fields such as surveying, navigation, and astrometry \cite{fiona2006keyboard}. Fiona et al. \cite{fiona2006keyboard} calculated the differences in arrival times of sound waves at two microphones installed at specific locations beside the keyboard. They employed both the maximum peak position approach and the correlation approach, achieving a recognition rate of up to 80\% with a 5-minute computation.
In Ranade's study \cite{ranade2009acoustic}, the triangulation method achieved an accuracy of 87.5\% in distinguishing between four keys on an ATM keypad.
In the paper by Bai et al. \cite{bai2021know}, they introduce a scheme that utilizes a single smartphone with dual microphones for conducting a side-channel attack. Their approach includes an efficient environment estimation scheme to overcome challenges related to microphone positioning variability and the scarcity of training data. The proposed scheme predicts keystrokes based on their locations and achieves a correct identification rate of 91.2\%.

\begin{figure}[h]
  \centering
  \includegraphics[width=0.4\textwidth]{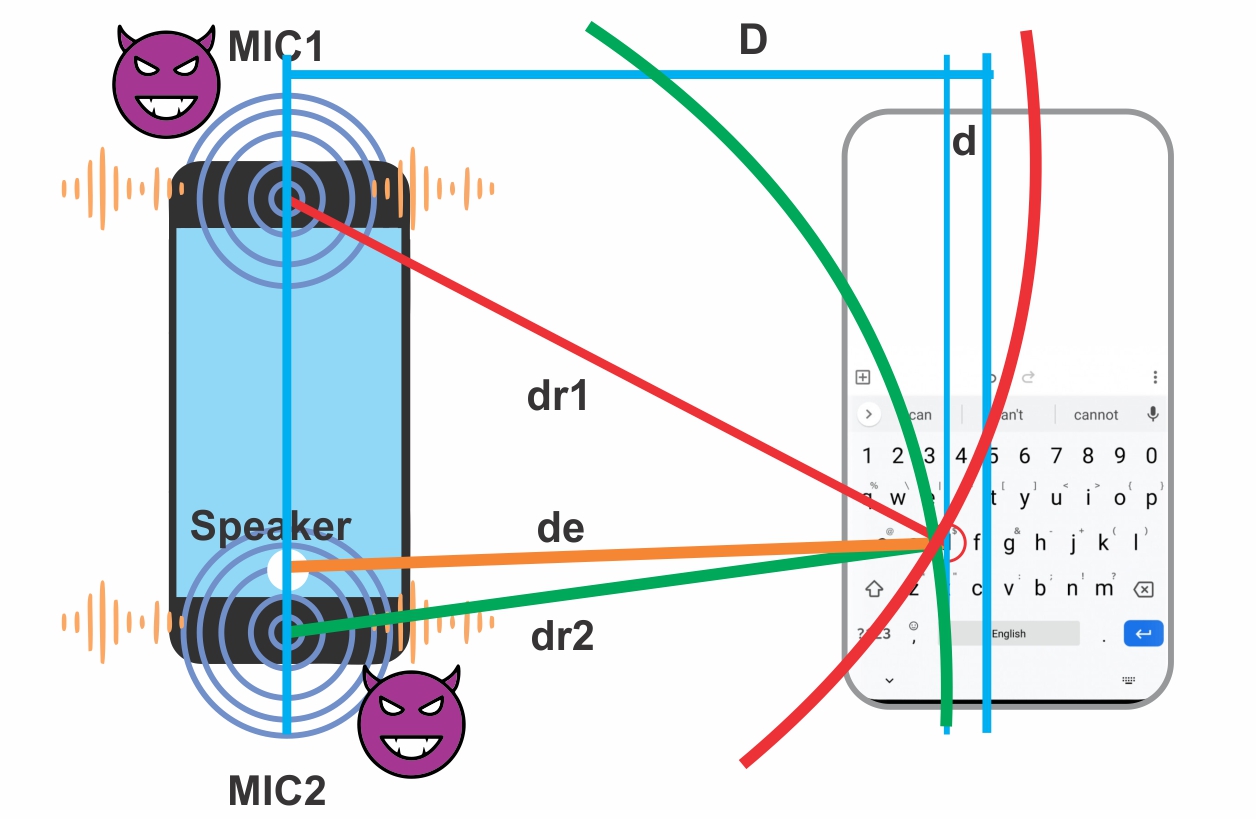}
  \caption{Foundational concepts of touchscreen keystroke localization using acoustic signal attenuation\cite{lu2019keylistener}.}
  \label{fig:keylistener}
\end{figure}

\subsubsection{Differential Audio Analysis}
\begin{figure}[h]
  \centering
  \includegraphics[width=0.4\textwidth]{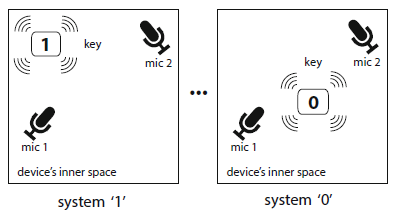}
  \caption{An ensemble involving the device's internal structure, key position, and two microphones creates a system necessitating transfer function estimation. The key classification relies on these derived transfer functions.\cite{de2019differential}.}
  \label{fig:Differential audio}
\end{figure}

In \cite{de2019differential}, the authors introduce a method called \textit{DAA} (differential audio analysis) for analyzing the differential attributes of sounds recorded by two microphones situated within the device's empty space. This analysis is conducted through the transfer function between the two captured signals. The method is employed on four-digit PIN pads. The authors successfully identify all 1200 keystrokes from two independently tested devices of the same model, yielding classification rates of 100\% and 99.8\% for the first two models and 63\% for the third model(Figure \ref{fig:Differential audio}).

\subsubsection{Phase Shifts and the Doppler Effect}
As shown in the \ref{fig:keylistener}, the study by Lu et al. \cite{lu2019keylistener} explores the reduction of acoustic signal strength, demonstrating that a user's typing fingers can be pinpointed by assessing the attenuation of acoustic signals captured by smartphone microphones. This attenuation is then harnessed to localize individual keystrokes and subsequently assess discrepancies arising from background noise. To enhance the precision of keystroke localization, \textit{KeyListener} additionally monitors finger movements during input through phase shifts and the Doppler effect, which aids in mitigating errors linked to the acoustic signal attenuation-based localization technique. Moreover, the application incorporates a binary tree-based search strategy to deduce keystrokes contextually. This approach achieves a near 90\% accuracy in inferring the correct keystroke among the top 5 candidates, accompanied by a top-5 error rate of approximately 6\%.

\section{Post Processing Technologies}
This section outlines methods commonly reliant on language models to improve keystroke identification quality. These methods contribute to the enhancement of word or sentence recognition rates. In fact, this method is used to filter incorrect classified characters to improve the recognition rate \cite{toreini2015acoustic}.

\subsection{Hidden Markov Models (\textit{HMMs})}
Hidden Markov Models (\textit{HMMs}) are a type of machine learning model. \textit{HMMs} is a statistical model used for modeling systems with hidden states that generate observable outcomes \cite{baum1966statistical,maleki}. In Fact, Hidden Markov models are engineered to handle data that can be represented as a ‘sequence’ of observations over time. They have applications in various fields, including speech recognition, natural language processing, bioinformatics, and more. \textit{HMMs} are used to model data sequences, where the underlying process is assumed to be a Markov process with hidden states that influence the observed data \cite{rabiner1989tutorial}. In the scenario of an acoustic side-channel attack on a keyboard, an \textit{HMM} is a model that undergoes training using a collection of text to anticipate the probable word or character at specific positions within a sequence. For example, if the result of an identification method is ‘clasc’, a \textit{HMM} could be used to infer that ‘c’ was, in fact, a falsely classified ‘s’ \cite{harrison2023practical}. In the work by Zhuang et al. \cite{zhuang2009keyboard}, the authors employ a variant of the Viterbi algorithm tailored for second-order \textit{HMMs}, resembling the approach outlined in Thede et al. \cite{thede1999second}. They have demonstrated that the language model correction greatly enhances the correct recovery rate for words. Additionally, they utilize the \textit{HMM} in an unsupervised manner for keyboard key recognition. This approach achieves an average recovery rate of 87.6\% for words and 95.7\% for characters.

\subsection{Spelling Checker}
Utilizing a spell checker represents a straightforward method to capitalize on language-related insights. In the study by Zhuang et al. \cite{zhuang2009keyboard}, they employed the Aspell \cite{atkinson2005aspell} spell check tool on identified text, and observed specific enhancements. Nevertheless, standard spell checkers have restrictions regarding the types of spelling errors they can address (for instance, a maximum of two incorrect letters in a word). Their functionality is optimized for typical errors made by human typists rather than the errors made by acoustic emanation classifiers. Hence, their effectiveness in this context is understandably restricted.

\subsection{N-GRAM Language Model}
The spelling correction approach discussed earlier doesn't consider word frequency relationships or grammatical concerns. For instance, certain words are more prevalent than others, and there are established rules for constructing sentences. The spelling correction process might approve "Ir fact" as accurate because "Ir" is a recognized word, even if the intended phrase is likely "In fact" \cite{zhuang2009keyboard}. A potential solution involves employing an n-gram language model that probabilistically represents word frequency and the associations between neighboring words \cite{jurafsky2003probabilistic}. To be precise, in \cite{zhuang2009keyboard}, trigrams are integrated with the aforementioned spelling correction technique, and sentences are structured using a graphical model.
\section{Methods and Results Comparison }
In this survey, we have delved into the feasibility and mechanics of acoustic side-channel attacks on keyboards, which stem from the sounds and vibrations generated during the act of typing. We explored the distinctive sounds produced by various keys, underpinned by the physics of keyboards and the diverse typing styles of users.

Throughout our analysis, we have highlighted the practicality of these attack scenarios, demonstrating that attackers can exploit not only physical proximity but also remote and indirect methods to record keystroke sounds. This poses a significant threat to the security and privacy of users, as attackers can capture sensitive information without direct physical access to the victim's device.

We conducted a thorough analysis of different methods used to launch acoustic side-channel attacks on keyboards. The presented comparison tables offer insights into the landscape of both timing-based and geometry-based attack strategies (Table \ref{table:1}), as well as signal processing, classical machine learning-based, and deep learning-based approaches (Table \ref{table:2}). We evaluated key features and performance metrics across various categories to gain insights into the diverse strategies used by researchers in this field.
It is evident that the choice of methodology and key features greatly influences the accuracy of these attacks, with some achieving remarkably high success rates. However, each approach also comes with its own set of limitations, such as dependencies on typing style, keyboard models, environmental noise, and the need for sizable training datasets. These are the typical constraints that generally affect most methods at their core:

\begin{itemize}
\item \textbf{Typing Style Dependency:} Regardless of the approach, the accuracy of acoustic side-channel attacks is inherently tied to the typing style of the user, making it challenging to generalize across different users.

\item \textbf{Keyboard Model Dependency:} All approaches exhibit a degree of dependency on the specific keyboard model being used, limiting their applicability to various keyboard types.

\item \textbf{Environmental Noise Sensitivity:} The presence of environmental noise can significantly impact the accuracy of these attacks, posing a common challenge across methodologies.

\item \textbf{Training Data Requirements:} Many of these methods require substantial training datasets, which may not always be readily available or feasible to collect.

\end{itemize}

The timing-based analysis involves using a range of techniques, such as hidden Markov models, cross-correlation, and distance metrics, to identify patterns. Usually, there are common limitations to these techniques, such as typing errors, ecological validity, and handling multiple identical time intervals.
Geometry-based approaches introduce innovative techniques like Time Difference of Arrival (TDOA) and acoustic triangulation, demonstrating commendable keystroke recovery rates, albeit with some restrictions regarding precise key positioning and acoustic signal source angles.

Signal processing methods, classical machine learning, and deep learning-based approaches emphasize the significance of feature extraction, including FFT coefficients, MFCC features, and spectrograms. While classical machine learning models like hidden Markov models and support vector machines exhibit robust performance, deep learning models, such as convolutional and recurrent neural networks, achieve promising accuracy rates, often surpassing 90\% in certain situations.
Furthermore, the limitations associated with each technique, such as ecological validity and recognition accuracy, should be carefully considered in real-world scenarios.

This survey equips researchers, practitioners, and security experts with a comprehensive understanding of the current state of acoustic side-channel attacks on keyboards. By illuminating the strengths, weaknesses, and potential challenges posed by these approaches, we aim to encourage further research and innovation in the realm of keyboard security. As the digital landscape continues to evolve, safeguarding against acoustic side-channel attacks remains a critical endeavor, and this survey serves as a resource in that pursuit.


\begin{table}[]
\resizebox{\columnwidth}{!}{%
\begin{tabular}{|cc|c|p{0.7in}|p{1.in}|p{1.in}|p{1.5in}|p{1.in}|}
\hline
\multicolumn{2}{|c|}{Approach} &
  Cite &
  Methodology &
  Key Features &
  Accuracy &
  Limitations &
  Attack Target Data \\ \hline
\multicolumn{2}{|c|}{\multirow{4}{*}{\rotatebox[origin=c]{90}{\parbox{35mm}{\centering Timing-based Analysis}}}} &
  \cite{foo2010timing} &
  Hidden Markov Model &
  The time interval between keystrokes &
  search space is reduced by at least two orders of magnitude &
  Multiple of the same time intervals &
  ATM and door keypads \\ \cline{3-8} 
\multicolumn{2}{|c|}{} &
  \cite{liu2019keystroke} &
  Cross-correlation &
  Inter-keystroke timing dictionary built from a human cognitive model &
  top-3, top-10, top-25, top-50, top-100 = 75\%, 80\%, 81\%, 83\%, 88\%&
  Typing styles, Typing errors, Ecological validity &
  PIN \\ \cline{3-8} 
\multicolumn{2}{|c|}{} &
  \cite{panda2020behavioral} &
  Cross-Correlation &
  Hold time, release time, and time interval between keystrokes &
  60\% chance to recover the PIN key &
  The size of the  training dataset &
  4–6 digit random PIN \\ \cline{3-8} 
\multicolumn{2}{|c|}{} &
  \cite{liu2019human} &
  Cosine similarity &
  Inter-keystroke timing dictionary &
  guessing the correct PIN is above 10\%. &
  Typing styles, Typing errors, Ecological validity &
  PIN \\ \hline
\multicolumn{2}{|c|}{\multirow{9}{*}{\rotatebox[origin=c]{90}{\parbox{55mm}{\centering Geometry-based approaches}}}} &
  \cite{zhu2014context} &
  \multirow{4}{*}{\rotatebox[origin=c]{90}{\parbox{30mm}{\centering Time Difference of Arrival}}} &
  \multirow{9}{*}{\rotatebox[origin=c]{90}{\parbox{75mm}{\centering Acoustic Signal}}} &
  recover 72.2\% of Keystrokes &
  Angle of Smartphones(sound recorder) &
  Keyboard \\ \cline{3-3} \cline{6-8} 
\multicolumn{2}{|c|}{} &
  \cite{rosmansyah2017microphone} &
   &
   &
  86\% accuracy &
  number of sound recorder devices &
  Keyboard \\ \cline{3-3} \cline{6-8} 
\multicolumn{2}{|c|}{} &
  \cite{liu2015snooping} &
   &
   &
  recover 94\% of keystrokes &
  the sampling rate, Microphone Placement &
  Keyboard \\ \cline{3-3} \cline{6-8} 
\multicolumn{2}{|c|}{} &
  \cite{cheng2022dictionary} &
   &
   &
  a 45.2\%success rate &
  Microphone's Placement, Word length &
  Keyboard \\ \cline{3-4} \cline{6-8} 
\multicolumn{2}{|c|}{} &
  \cite{fiona2006keyboard} &
  \multirow{3}{*}{\rotatebox[origin=c]{90}{\parbox{20mm}{\centering Acoustic Triangulation}}} &
   &
  80\%accuracy &
  cannot make precise judgments on the positions of close keys &
  Keyboard \\ \cline{3-3} \cline{6-8} 
\multicolumn{2}{|c|}{} &
  \cite{ranade2009acoustic} &
   &
   &
  87.5\% accuracy &
  Microphone resolution &
  ATM keypad \\ \cline{3-3} \cline{6-8} 
\multicolumn{2}{|c|}{} &
  \cite{bai2021know} &
   &
   &
  correctly identify 91.2\% of keystrokes with 10-fold cross-validation &
  keyboard model &
  keyboard text \\ \cline{3-4} \cline{6-8} 
\multicolumn{2}{|c|}{} &
  \cite{de2019differential} &
  Differential Audio Analysis &
   &
  A classification rate of 99.8\%. &
  Non-audible Pin pads &
  PIN pads \\ \cline{3-4} \cline{6-8} 
\multicolumn{2}{|c|}{} &
  \cite{lu2019keylistener} &
  Phase Shifts and the Doppler Effect &
   &
  90\% on a top-5 &
  Microphone's placement &
  keyboards of touch screen \\ \hline
\end{tabular}
}
\caption{Comparison of Timing-Based and Geometry-Based Acoustic Side Channel Attack Approaches on Keyboards}
\label{table:1}
\end{table}
\restoregeometry

\newpage

\begin{table}[]
\resizebox{\columnwidth}{!}{%
\begin{tabular}
{|c|c|p{1.0in}|p{1.5in}|p{1.in}|p{1.5in}|p{1.5in}|} 
\hline
\multicolumn{1}{|l|}{Appr.} &
  Cite &
  Methodology &
  Key Features &
  Accuracy &
  Limitations &
  Attack Target Data \\ \hline
\rotatebox[origin=c]{90}{\parbox{25mm}{\centering Signal Processing}} &
  \cite{ponnam2013keyboard} &
  Distance Metric Approach &
  FFT and raw audio signals &
  60\% recognition rate &
  Typing styles, Keyboard model changing &
  Password \\ \hline
\multirow{10}{*}{\rotatebox[origin=c]{90}{\parbox{90mm}{\centering Classical Machine Learning-based}}} &
  \cite{zhuang2009keyboard} &
  Hidden Markov Model &
  cepstrum features &
  96\%, 90\%, 80\%&
  Constraints of English language &
  English text and random passwords on keyboard \\ \cline{2-7} 
 &
  \cite{wit2014all} &
  K-means &
  coefficients of the Mel-frequency Cepstrum &
  90\%&
  Typing speed, Ambient noise &
  keyboard text \\ \cline{2-7} 
 &
  \cite{wang2016accurate} &
  Support Vector Machines &
  Blind Source Separation (BSS) and Independent Component Analysis (ICA), and MFCC features &
  78.4\% recognition accuracy &
  Microphone's placement &
  Two combined keystrokes \\ \cline{2-7} 
 &
  \cite{anand2018keyboard} &
  Logistic Regression, Random Forest, Linear Nearest Neighbor Search &
  FFT coefficients and MFCC &
  accuracy of 74.33\%&
  Noisy Audio Signals &
  Random passwords, PINs \\ \cline{2-7} 
 &
  \cite{qin2019lol} &
  Support Vector Machines anomaly detection and Back Propagation Neural Network &
  spectrum features &
  99.47\% keystroke detection rate, a 97.27\% recognition accuracy and 84.55\% content recovery accuracy &
  Combined keys &
  keyboard text \\ \cline{2-7} 
 &
  \cite{halevi2012closer} &
  Time-frequency decoding &
  MFCC features &
  detecting correctly the typed key is 64\% per character &
  Typing style &
  Random passwords \\ \cline{2-7} 
 &
  \cite{asonov2004keyboard} &
  JavaNNS neural network &
  time-FFT &
  recognized correctly 79\% of clicks &
  Silent keyboards, touchscreen or touchstream &
  ATM keypads \\ \cline{2-7} 
 &
  \cite{berger2006dictionary} &
  Cross-correlation &
  FFT Coefficients &
  90\% of finding the correct word &
  Sound duration, word length, repeated characters, Shift keys, punctuation marks and digits &
  7-13 characters from typing on a keyboard \\ \cline{2-7} 
 &
  \cite{compagno2017don} &
  Logistic Regression &
  Mel-frequency cepstral coefficients (MFCC) &
  top-5 accuracy of 91.7\%&
  Typing style and keyboard model &
  random key pressed on keyboard \\ \cline{2-7} 
 &
  \cite{cecconello2019skype} &
  Logistic Regression &
  MFCC &
  top-5 accuracy of 91.7\%&
  keyboard model &
  random key pressed on keyboard \\ \hline
\multirow{7}{*}{\rotatebox[origin=c]{90}{\parbox{50mm}{\centering Deep Learning-based}}} &
  \cite{harrison2023practical} &
  Self-attention transformer &
  Mel-Spectrograms &
  95\% accuracy &
  typing style and randomized passwords &
  laptop keyboard \\ \cline{2-7} 
 &
  \cite{giallanza2019keyboard} &
  Convolutional and recurrent neural networks &
  Cepstrum &
  ability to detect 41.8\% of keystrokes and 27\% of words &
  Microphone's placement, and keyboard models &
  Keyboard \\ \cline{2-7} 
 &
  \cite{akinbi4431909password} &
  Vision transformer based (ConvMix) &
  Spectrogram &
  92.44\% accuracy &
  The number of samples &
  Passwords \\ \cline{2-7} 
 &
  \cite{toreini2015acoustic} &
  Neural Networks &
  MFCC coeficients &
  success rate of 84\%&
  - &
  Enigma keyboard \\ \cline{2-7} 
 &
  \cite{kelly2010cracking} &
  Hidden Markov Model and neural network newpnn() and Spellchecker &
  FFT and Cepstrum &
  81.47\%&
  lack of publically dataset &
  Keystroke sequence \\ \cline{2-7} 
 &
  \cite{slater2019robust} &
  Recurrent neural network &
  FFT &
  92.60\%&
  keyboard Model, Environmental noise and microphone placement &
  Keyboard text \\ \cline{2-7} 
 &
  \cite{martinasek2015acoustic} &
  Multi-layer perceptron &
  Spectrogram &
  72.3\% success rate &
  Size of the spectrogram matrix &
  Keyboard \\ \hline
\end{tabular}
}
\caption{Comparison of Frequency-Based Acoustic Side Channel Attack Approaches on Keyboards}
\label{table:2}
\end{table}

\restoregeometry 

\printbibliography
\end{document}